\documentclass[preprint,showpacs,titlepage,aps,prd,epsf,epsfig,
tightenlines,amsmath,byrevtex,nofootinbib,subfigure]{revtex4}

\usepackage{graphicx,subfigure}

\def\lsim{\raise0.3ex\hbox{$\;<$\kern-0.75em\raise-1.1ex
\hbox{$\sim\;$}}}
\def\gsim{\raise0.3ex\hbox{$\;>$\kern-0.75em\raise-1.1ex
\hbox{$\sim\;$}}}

\begin{document}

\title{Probing Nonstandard Neutrino Physics by 
Two Identical Detectors with Different Baselines}

\author{Nei Cipriano Ribeiro}
\altaffiliation{
Also at: 
Centro de Educa\c{c}\~ao Tecnol\'ogica de Campos, 
Campos dos Goytacazes, 28030-130, RJ, Brazil
}
\affiliation{
Departamento de F\'{\i}sica, Pontif{\'\i}cia Universidade Cat{\'o}lica
do Rio de Janeiro, C. P. 38071, 22452-970, Rio de Janeiro, Brazil} 

\author{Takaaki Kajita}
\altaffiliation{
Also at: 
Institute for the Physics and Mathematics of the Universe (IPMU),
University of Tokyo, Kashiwa, Chiba 277-8582, Japan
}
\affiliation{
Research Center for Cosmic Neutrinos, Institute for Cosmic Ray Research (ICRR),
University of Tokyo, Kashiwa, Chiba 277-8582, Japan}

\author{Pyungwon Ko}
\affiliation{
School of Physics, Korean Institute for Advanced Study, Seoul 130-722, Korea} 

\author{Hisakazu Minakata}
\affiliation{
Department of Physics, Tokyo Metropolitan University, \\ Hachioji,
Tokyo 192-0397, Japan}

\author{Shoei Nakayama}
\altaffiliation{
Current address: 
Kamioka Observatory, Institute for Cosmic Ray Research,
University of Tokyo, Kamioka, Hida, Gifu 506-1205, Japan
}
\affiliation{
Research Center for Cosmic Neutrinos, Institute for Cosmic Ray Research (ICRR),
University of Tokyo, Kashiwa, Chiba 277-8582, Japan}

\author{Hiroshi Nunokawa}
\affiliation{
Departamento de F\'{\i}sica, Pontif{\'\i}cia Universidade Cat{\'o}lica
do Rio de Janeiro, C. P. 38071, 22452-970, Rio de Janeiro, Brazil}

\date{\today}

\begin{abstract}
The Kamioka-Korea two detector system is a powerful experimental
setup for resolving neutrino parameter degeneracies and probing CP
violation in neutrino oscillation.
In this paper, we study sensitivities of this same setup
to several nonstandard neutrino physics such as quantum decoherence,
tiny violation of Lorentz symmetry, and nonstandard interactions of
neutrinos with matter. In most cases, the Kamioka-Korea two-detector setup
is more sensitive than the one-detector setup, except for the Lorentz
symmetry violation with CPT violation, and the nonstandard neutrino
interactions with matter. It can achieve significant improvement
on the current bounds on nonstandard neutrino physics.
\end{abstract}
\vglue 0.6cm


\maketitle


\section{Introduction}
\label{introduction}

Variety of the neutrino experiments, the atmospheric \cite{SKatm},
solar \cite{solar}, reactor \cite{KamLAND},
and accelerator \cite{accelerator} experiments, have been successful
in identifying the neutrino mass induced neutrino oscillation as a
dominant mechanism for neutrino disappearances.
After passing through the discovery era,
the neutrino physics will enter the epoch of precision study,
as the CKM phenomenology and a detailed study of CP violation
have blossomed in the quark sector.
The MNS (Maki-Nakagawa-Sakata) matrix elements 
 will be measured with higher accuracy,
including the CP phase(s), and the neutrino properties such as their
interactions with matter etc., will be studied in a greater accuracy.
During the course of precision studies, it will become natural
to investigate nonstandard physics related with neutrinos, which
include flavor changing neutral/charged current
interactions \cite{wolfenstein,NSI,grossmann},
the effects of quantum decoherence \cite{ellis,QD-bari1,benatti-floreanini}
that  may possibly arise due to quantum gravity at short
distance scale \cite{ellis}, and 
violation of Lorentz and CPT invariance \cite{Coleman1,Coleman2,Kostelecky},
to name a few.

It is well known that in history of physics experiments based on interference
effect played very crucial roles in advancing our understanding of the
physical laws.
The famous two-slit experiment by Young,
Michelson-Morley experiment which demonstrated that
there is no ether, Davidson-Germer experiment on electron diffraction,
$K^0 - \overline{K^0}$ oscillation, etc.
Likewise, neutrino oscillation experiments may probe another important
structure of fundamental physics  by observing tiny effects due to
CPT violation or quantum decoherence, which may be rooted in
quantum gravity. Along with the neutral meson systems
($K^0 - \overline{K^0}$ and $B^0 - \overline{B^0}$),
neutrino oscillations could provide competing
and/or complementary informations on those exotic  effects.
See Ref.~\cite{Mavromatos:2006yn} for a recent review on this subject.

In a previous work \cite{T2KK1st,T2KK2nd} we have introduced and
explored in detail the physics potential of the Kamioka-Korea two detector
setting which receive an intense neutrino beam from ~{J-PARC}.
We have demonstrated that the setting is powerful enough to
resolve all the eight-fold parameter degeneracy
\cite{intrinsic,MNjhep01,octant},
if $\theta_{13}$ is in reach of the next generation accelerator
\cite{T2K,NOVA} and the reactor experiments \cite{MSYIS,reactor13}.
The degeneracy includes the parameters $\theta_{13}$, $\delta$ and
octant of $\theta_{23}$, and it is doubled by the ambiguity which
arises due to the unknown sign of $\Delta m^2_{31}$.
The detector in Korea plays a decisive role to lift the last one.
For related works on Kamioka-Korea two detector complex, see, for example,
\cite{hagiwara,Okumura-Seoul2006,Dufour-Seoul2006,Rubbia-Seoul2006}.

The Kamioka-Korea identical two detector setting is a unique
apparatus for studying nonstandard physics (NSP).
As will be elaborated in Sec.~\ref{NSP} the deviation from the
expectation by the standard mass-induced oscillation can be
sensitively probed by comparing yields at the intermediate (Kamioka)
and the far (Korea) detectors.
In this paper, we aim at exploring physics potential of the Kamioka-Korea
setting in a systematic way.
By this we mean that we examine several NSP effects in a single
framework by concentrating on $\nu_\mu - \nu_\tau$ subsystem
in the standard three-flavor mixing scheme, and focus on $\nu_\mu$
disappearance measurement. Though limited framework,
it will allow us to treat the problem in a coherent fashion.

In analyzing nonstandard physics in this paper,
we aim at demonstrating the powerfulness of the Kamioka-Korea
identical two detector setting, compared to other settings.
For this purpose, we systematically compare the results obtained with
the following three settings: 

\begin{itemize}

\item

Kamioka-Korea setting:
Two identical detectors one at Kamioka and the other in Korea each 0.27 Mton

\item
Kamioka-only setting: A single 0.54 Mton detector at Kamioka

\item
Korea-only setting: A single 0.54 Mton detector at somewhere in Korea.

\end{itemize}
Among the cases we have examined Kamioka-Korea setting always
gives the best sensitivities, apart from two exceptions of violation
of Lorentz invariance in a CPT violating manner, and the nonstandard
neutrino interactions with matter.
Whereas, the next best case is sometimes Kamioka-only or Korea-only settings
depending upon the problem.

This paper is organized as follows.
In Sec.~\ref{NSP}, we illustrate how we can probe nonstandard physics
with Kamioka-Korea two detector setting, with a quantum decoherence
as an example of nonstandard neutrino physics.
In Sec.~\ref{method}, we describe the statistical method which is used
in our analyses in the following sections.
In Sec.~\ref{decoherence}, we discuss quantum decoherence.
In Sec.~\ref{lorentz}, we discuss possible violation of Lorentz invariance.
In Sec.~\ref{NSI}, we discuss non-standard neutrino matter interactions,
and the results of study is summarized in Sec.~\ref{conclusion}.


\section{Probing non-standard physics with Kamioka-Korea Two Detector Setting}
\label{NSP}

\begin{figure}
\vspace{-60mm}
\centering
\includegraphics[width=16.0cm]{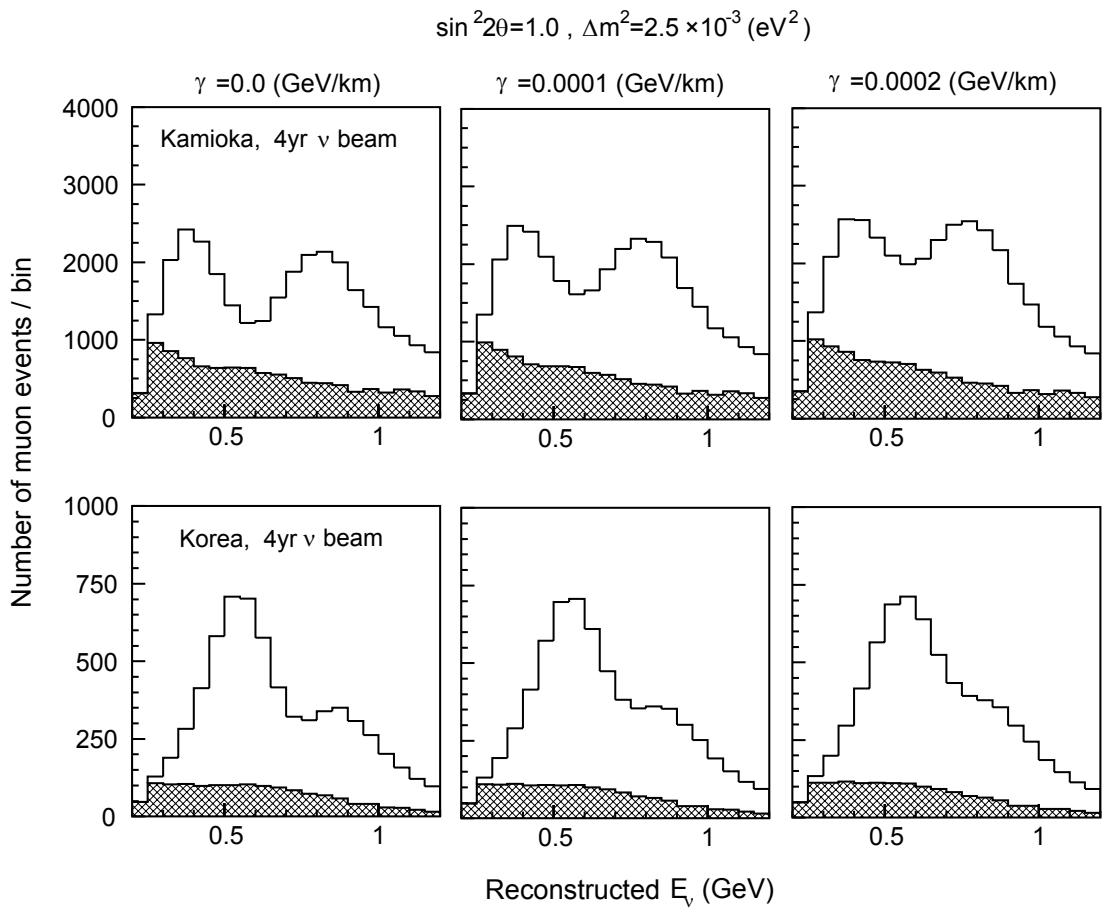}
\vspace{-40mm}
\caption{\label{example_event}
Event spectra of neutrinos at Kamioka (the top panel) and Korea
(the bottom panel)
for $\gamma = 0$ (the left column), $1 \times 10^{-4}$ GeV/km 
(the middle column), and $\gamma = 2 \times 10^{-4}$ GeV/km 
(the right column). The hatched areas denote the contributions from 
non-quasi-elastic events.}
\end{figure}

In this section, we describe how we proceed in the following sections.
For the purpose of illustration, we consider quantum decoherence (QD)
as nonstandard neutrino physics.
In this case, the $\nu_\mu$ survival probability
(and the $\overline{\nu}_\mu$ survival probability assuming CPT invariance
in the presence of QD) is given by~\cite{QD-bari1,benatti-floreanini}, 
\begin{equation}
P ( \nu_\mu \rightarrow \nu_\mu ) = P ( \overline{\nu}_\mu \rightarrow
\overline{\nu}_\mu ) =
1 - {1\over 2}~\sin^2 2 \theta ~
\left[ 1 - e^{-\gamma(E)L} \cos \left( {\Delta m^2 L \over 2 E}
\right) \right],
\end{equation}
where we consider the case $\gamma(E) = \gamma / E$
(see Sec. IV where we consider also the cases of 
$\gamma(E)$ which has other energy dependences). 
Then one can calculate the number of $\nu_\mu$ and $\overline{\nu}_\mu$
events observed at two detectors placed at Kamioka (with baseline of 295 km)
and Korea (with baseline of 1050 km),
using the above survival probability and the neutrino beam profiles.
For simplicity, let us consider the number of observed neutrino events
both at Kamioka and Korea, for each energy bin (with 50 MeV width) from
$E_\nu = 0.2$ GeV upto $E_\nu = 1.4$ GeV.
In Fig.~\ref{example_event}, we show the $\nu_\mu$ event spectra
at detectors located at Kamioka and Korea for the pure oscillation 
$\gamma = 0$ (the left column) and the oscillation plus QD with two 
different QD parameters, $\gamma = 1 \times 10^{-4}$ GeV/km (the middle 
column) and  $\gamma = 2 \times 10^{-4}$ GeV/km 
(the right column)
\footnote{In order to convert this $\gamma$ in unit of GeV/km to
$\gamma$ defined in Eq.~(5), one has to multiply $0.197 \times 10^{-18}$.}.
One finds that spectra change for non-vanishing $\gamma$. Especially we 
point out that the spectra changes are different between detectors at 
Kamioka and Korea due to the different $L/E$ values at the two positions.

Assuming the actual data at Kamioka and Korea are given (or well described) 
by the pure oscillation with $\sin^2 2 \theta = 1$ and
$\Delta m^2 = 2.5 \times 10^{-2}$ eV$^2$, we could claim that
$\gamma = 1 \times 10^{-4}$ GeV/km (shown in the middle column),
for example, would be inconsistent with the data. One can make this kind of
claim in a more proper and quantitative manner using the $\chi^2$ analysis,
which is described in details in the following section.

\section{Analysis Method}
\label{method}

In studying nonstandard physics through neutrino oscillations
we restrict ourselves into the $\nu_\mu - \nu_\tau$ subsystem,
rather than dealing with the full three generation problem.
The reason is partly technical and partly physics motivated.
The technical reason for truncation is to simplify analysis in a
manner not to spoil the most important features of the problem.
First of all, $\nu_\mu \to \nu_\mu$ and $\bar{\nu}_\mu \to \bar{\nu}_\mu$
probabilities do not depend much on the yet unknown paramters, namely, 
$\theta_{13}$, CP phase and the sign of $\Delta m^2_{31}$.
Second, in most cases the earth matter effect is a sub-leading
effect in the $\nu_\mu - \nu_\tau$ subsystem 
so that by restricting to the subsystem we need not to worry
about complications due to the matter effect. 
Moreover, many of the foregoing analyses were carried out
under the truncated framework.
By working under the same approximation the comparison between
ours and the existing results becomes much simpler.
Thus, the $\Delta m^2$ which will appear in neutrino oscillation
probabilities in the following sections is meant to be
$\Delta m^2 \equiv \Delta m^2_{32} = m^2_{3} - m^2_{2} $.
$\theta$ will be a mixing angle in the unitary matrix diagonalizing
the Hamiltonian in the $\nu_\mu - \nu_\tau$ subsystem, which is
essentially equal to $\theta_{23}$.

\subsection{Method of statistical analysis}
\label{statistical}

In order to understand the sensitivity of the experiment
with the two detector system at
295~km (Kamioka) and 1050~km (Korea), we carry out
a $\chi^2$ analysis.
In the present analysis, we only included $\nu_{\mu}$ and
$\overline{\nu}_{\mu}$ disappearance channels.
In short, the definition of the statistical procedure is similar to the
one used in \cite{T2KK2nd} excluding the
electron events.
The assumption on the experimental setting is also identical
to that of the best performance setting identified in Ref.~\cite{T2KK1st}.
Namely, 0.27~Mton fiducial masses for the intermediate
site (Kamioka, 295~km) and the far site (Korea, 1050~km).
For the reference, we also consider 0.54~Mton detector for Kamioka
or Korea only.
The neutrino beam is assumed to be 2.5 degree off-axis one produced
by the upgraded J-PARC 4~MW proton beam.
It is assumed that the experiment will continue for 8 years
with 4 years of neutrino and 4 years of anti-neutrino runs.

We use various numbers and distributions available from
references related to T2K  \cite{JPARC-detail}, in which many of
the numbers are updated after the original proposal \cite{T2K}.
Here, we summarize the main assumptions and the methods
used in the $\chi^2$ analysis.
We use the reconstructed neutrino energy for single-Cherenkov-ring
muon events.
The resolution in the reconstructed neutrino energy is 80~MeV
for quasi-elastic (QE) events.
We take
$\Delta m^2 = 2.5 \times 10^{-3} \text{eV}^2$
and $\sin^2 2 \theta_{23} = 1.0 $ for
our reference value. However, whenever we expect that there is
a correlation between the expected sensitivity to new physics and the
oscillation parameters, we scan
$\Delta m_{31}^2$ between 2.0 and 3.0$\times 10^{-3}$eV$^2$
and $\sin^2 2 \theta_{23}$ between 0.9 and 1.0.
The shape of the
energy spectrum for the anti-neutrino beam is assumed to be
identical to that of the neutrino beam. The event rate
for the anti-neutrino beam
in the absence of neutrino oscillations
is smaller by a factor of 3.4 due mostly to the lower neutrino
interaction cross sections and partly to the slightly lower
flux. In addition, the contamination of the wrong sign
muon events is higher in the anti-neutrino beam.

We stress that in the present setting
the detectors located in Kamioka and in Korea are not only
identical but also receive neutrino beams with essentially
the same energy distribution (due to the same off-axis angle of 2.5 degree)
in the absence of oscillations.
However, it was realized recently that, due to a non-circular shape
of the decay pipe of the J-PARC neutrino beam line,
the flux energy spectra viewed at detectors in Kamioka and in
Korea are expected to be slightly different even at the same off-axis angle,
especially in the high-energy tail of the spectrum
\cite{Rubbia-Meregaglia-Seoul2006}.
The possible difference between fluxes in the intermediate and the
far detectors is taken into account as a systematic error in
the present analysis.

We compute neutrino oscillation probabilities
by numerically integrating neutrino evolution equation
under the constant density approximation.
The average density is assumed to be 2.3 and 2.8 g/cm$^3$ for the
matter along the beam line between the production target and
Kamioka and between the target and Korea, respectively \cite{T2KK1st}.
We assume that the number of electron with respect to that of nucleons
to be 0.5 to convert the matter density to the electron number density.

The statistical significance of the measurement considered in this
paper was estimated by using the following
definition of $\chi^2$:
\begin{eqnarray}
\chi^2 =  \sum_{k=1}^{4} \left(
\sum_{i=1}^{20}
\frac{\left(N(\mu)_{i}^{\rm obs} - N(\mu)_{i}^{\rm exp}\right)^2}
{ \sigma^2_{i} }
\right)
+ \sum_{j=1}^{4} \left(\frac{\epsilon_j}
{\tilde{\sigma}_{j}}\right)^2,
\label{equation:chi2def}
\end{eqnarray}
where
\begin{eqnarray}
 N(\mu)_{i}^{\rm exp} = N_{i}^{\rm non-QE} \cdot
   (1+\sum_{j=1,3,4} f(\mu)_{j}^{i}\cdot\epsilon_{j})
    + N_i^{\rm QE} \cdot
   (1+\sum_{j=1,2,4}   f(\mu)_{j}^{i}\cdot\epsilon_{j}) ~.
\label{equation:mu-number}
\end{eqnarray}

\noindent
In Eq.~(\ref{equation:chi2def}), 
$N(\mu)^{\rm obs}_i$ is the number single-ring muon events to be
observed for the given (oscillation) parameter set,
and $N(\mu )^{\rm exp}_i$ is the expected number of
events for the assumed parameters
in the $\chi^2$ analysis.
$k=1,2,3$ and $4$ correspond to the four combinations
of the detectors in Kamioka and in Korea with the
neutrino and anti-neutrino beams,
respectively.
The index $i$ represents the reconstructed neutrino energy bin
for muons.
The energy range for the muon events covers
from 200 to 1200~MeV. Each energy bin has 50~MeV width.
$\sigma_i$ denotes the
statistical uncertainties in the expected data.
The second term
in the $\chi^2$ definition collects the
contributions from variables which parameterize the systematic
uncertainties in the expected number of signal and background events.

$N^{\rm non-QE}_i$ are the number of non-quasi-elastic muon events
for the $i^{\rm th}$ bin 
whereas $N^{\rm QE}_i$ are the number of quasi-elastic muon events.
We treat the non-quasi-elastic and quasi-elastic muon events separately,
since the neutrino energy cannot be properly reconstructed for
non-quasi-elastic events.
Both $N^{\rm QE}_i$ and $N^{\rm non-QE}_i$ depend on
neutrino (oscillation) parameters but in a different way,
namely, the former (latter) being affected in direct (indirect) 
manner and hence dependence is strong (weak) as we can see
from the solid histgram ($N^{\rm QE}_i$ + $N^{\rm non-QE}_i$)
and the hatched region ($N^{\rm non-QE}_i$) in Fig.~\ref{example_event}. 
The key to high sensitivity to NSP is the different oscillation 
parameter dependence between Kamioka and the Korean detectors 
due to different baselines. 
The uncertainties in  $N^{\rm non-QE}_i$ and $N^{\rm QE}_i$
are represented by 4 parameters $\epsilon_j$ ($j=1$ to 4).

During the fit, the values of $N(\mu)^{\rm exp}_i$ are recalculated
for each choice of the (oscillation) parameters which are varied freely
to minimize $\chi^2$, and so are the systematic error parameters
$\epsilon_j$.
The parameter $f(\mu)^i_j$ represents
the fractional change in the predicted
event rate in the $i^{\rm th}$ bin due to a variation of the parameter
$\epsilon_j$.
  We assume that the experiment is equipped with a near detector
which measures the un-oscillated muon spectrum.
The uncertainties in the absolute normalization of events
are assumed to be 5\%
($\tilde{\sigma}_{1}$=0.05).
      The functional form of
      $f(\mu)_2^i = (E_\nu(rec)-800~\text{MeV}) / 800~\text{MeV}$
      is used to define the uncertainty in the spectrum shape for
      quasi-elastic muon events ($\tilde{\sigma}_{2}$=0.05) \cite{T2KK2nd}.

The uncertainty in the separation of quasi-elastic and
non-quasi-elastic interactions in the muon events is
assumed to be 20\% ($\tilde{\sigma}_{3}$=0.20).
In addition, for the number of events in Korea, the possible
flux difference between Kamioka and Korea is taken into account
in $f(\mu)_4^i$. The predicted flux
difference \cite{Rubbia-Meregaglia-Seoul2006}
is simply assumed to be the 1~$\sigma$ uncertainty in the flux
difference ($\tilde{\sigma}_{4}$).

Finally, in this work, the sensitivity at 90\% (99 \%) confidence 
level (CL) is defined by
\[
\Delta \chi^2 \equiv \chi_{\text{min}}^2 ({\rm osc.+nonstandard~ physics}) 
- \chi_{\text{min}}^2 ({\rm osc.}) \geq
2.71 \ (6.63),
\]
corresponding to the one degree of freedom. 
Similarly, the criterion for the two degrees of freedom is 
$\Delta \chi^2 \geq 4.61 \ (9.21)$.  


\section{Quantum decoherence (QD)}
\label{decoherence}

Study of ``quantum decoherence'' is based on a hypothesis
that somehow there may be a loss of coherence due to environmental
effect or quantum gravity and space-time foam, etc~\cite{ellis}.
We do not discuss the origin of decoherence in this paper, but
concentrate on how this effect can be probed by the Kamioka-Korea setting
(this statement also applies to other nonstandard physics 
considered in this paper).
For previous analyses of decoherence in neutrino experiments, see e.g.,
\cite{QD-bari1,gago,QD-bari2,hooper,barenboim}.
As discussed in Sec.~\ref{introduction} we consider the
$\nu_\mu - \nu_\tau$ two-flavor system.
Since the matter effect is a sub-leading effect in this channel
we employ vacuum oscillation approximation in this section.
The two-level system in vacuum in the presence of quantum decoherence
can be solved to give the $\nu_{\mu}$ survival probability
\cite{QD-bari1,benatti-floreanini}:
\begin{eqnarray}
P(\nu_{\mu} \rightarrow \nu_{\mu}) =
1 -  \frac{1}{2} \sin^2 2 \theta
\left[ 1 - e^{- \gamma (E) L}
\cos \left( \frac{ \Delta m^2 L} { 2 E} \right)
\right],
\label{Pmumu}
\end{eqnarray}
where $\gamma (E) > 0$ is the parameter which controls the strength of
decoherence effect.
Notice that the conventional two-flavor oscillation formula
is reproduced in the limit $\gamma (E) \rightarrow 0$.
Since the total probability is still conserved in the presence of QD,
the relation
$P(\nu_{\mu} \rightarrow \nu_{\tau}) = 1- P(\nu_{\mu} \rightarrow \nu_{\mu})$
holds.
A similar expression holds for the anti-neutrino survival probability
with possible different decoherence index $\bar{\gamma} (E)$.
We assume CPT invariance in this section so that
$\bar{\gamma} (E) = \gamma(E)$; Then, the $\bar{\nu}_{\mu}$
survival probability is the same as in (\ref{Pmumu}) follows.

Unfortunately, nothing is known about the energy dependence of $\gamma (E)$.
Therefore, we examine, following \cite{QD-bari1},
several typical cases of energy dependence of $\gamma (E)$:
\begin{eqnarray}
\gamma (E) = \gamma
\left( \frac{ E }{ \text{ GeV } } \right)^n ~({\rm with}~ n=0,2,-1)
\label{E-dep}
\end{eqnarray}
In this convention, the overall constant $\gamma$ has a dimension of
energy or (length)$^{-1}$, irrespective of the values of the exponent $n$.
We will use $\gamma$ in GeV unit in this section.
In the following three subsections, we analyze three different
energy dependences, $n=0, -1, 2$ one by one.

\subsection{Case of $\frac{1}{E}$ dependence of $\gamma (E)$ }

\begin{figure}[bhtp]
\begin{center}
\includegraphics[height=14cm]{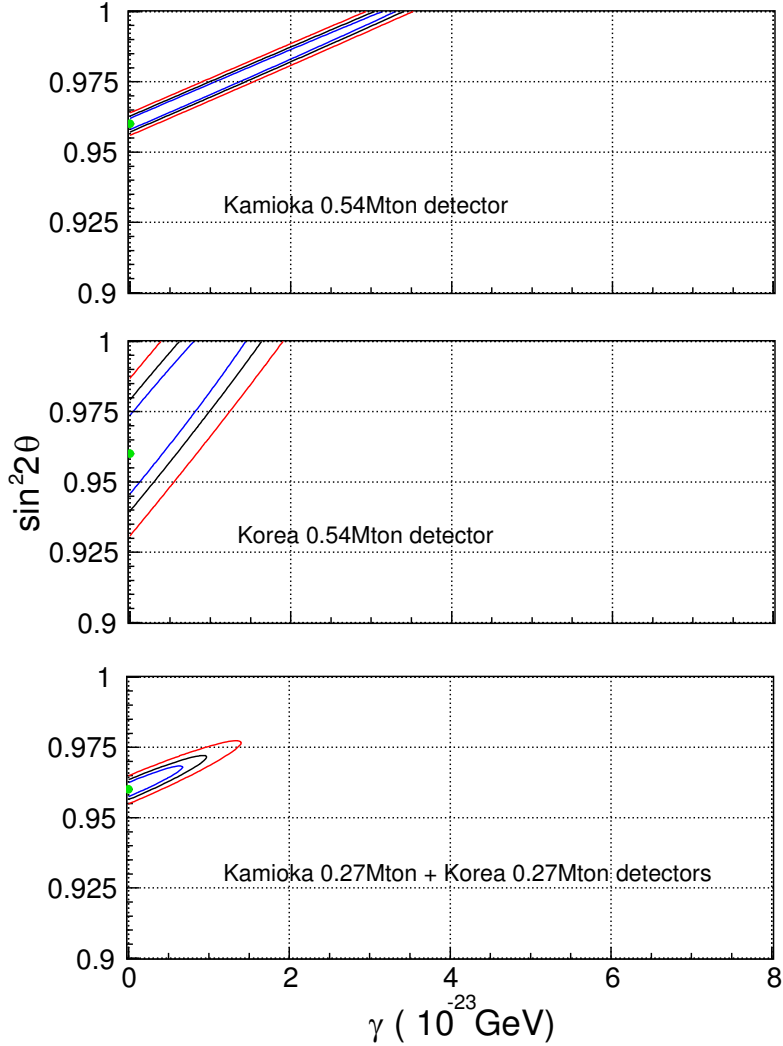}
\end{center}
\vglue -0.5cm
\caption{
The correlations between $\gamma$ and $\sin^2 2 \theta$ for three
experimental setups we consider: Kamioka-only, Korea-only and Kamiok-Korea.
Blue, black and red curves represent the contours for
68\% , 90\% and 99\% CL for two degrees of freedom.
Input values are $\gamma=0$, $\Delta m^2=2.5\times 10^{-3}$ eV$^2$ and 
$\sin^2 2 \theta=0.96$.
}
\label{sens-examples}
\end{figure}

\begin{figure}[bhtp]
\begin{center}
\hglue -0.5cm
\includegraphics[width=0.50\textwidth]{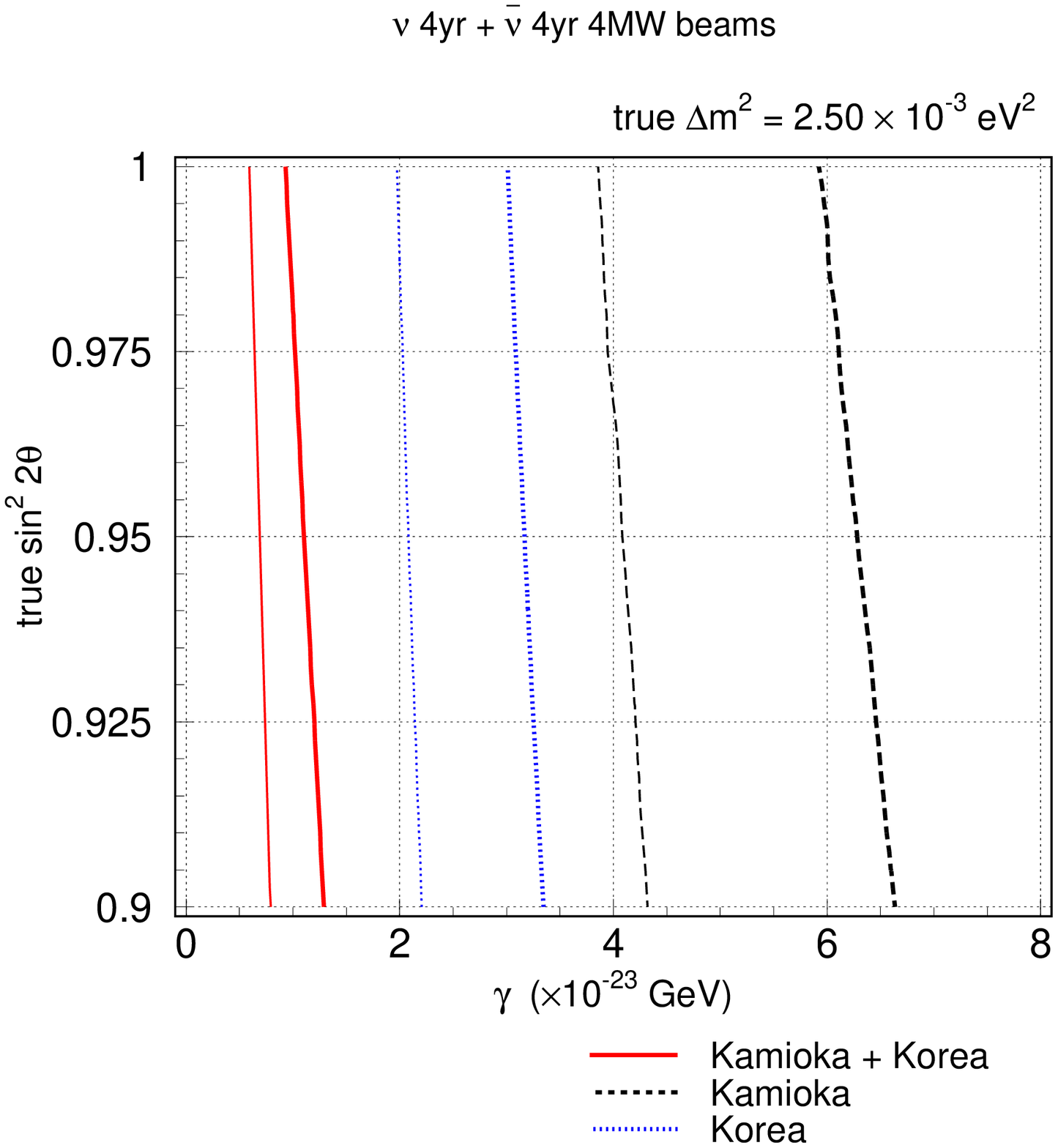}
\hglue  0.2cm
\includegraphics[width=0.50\textwidth]{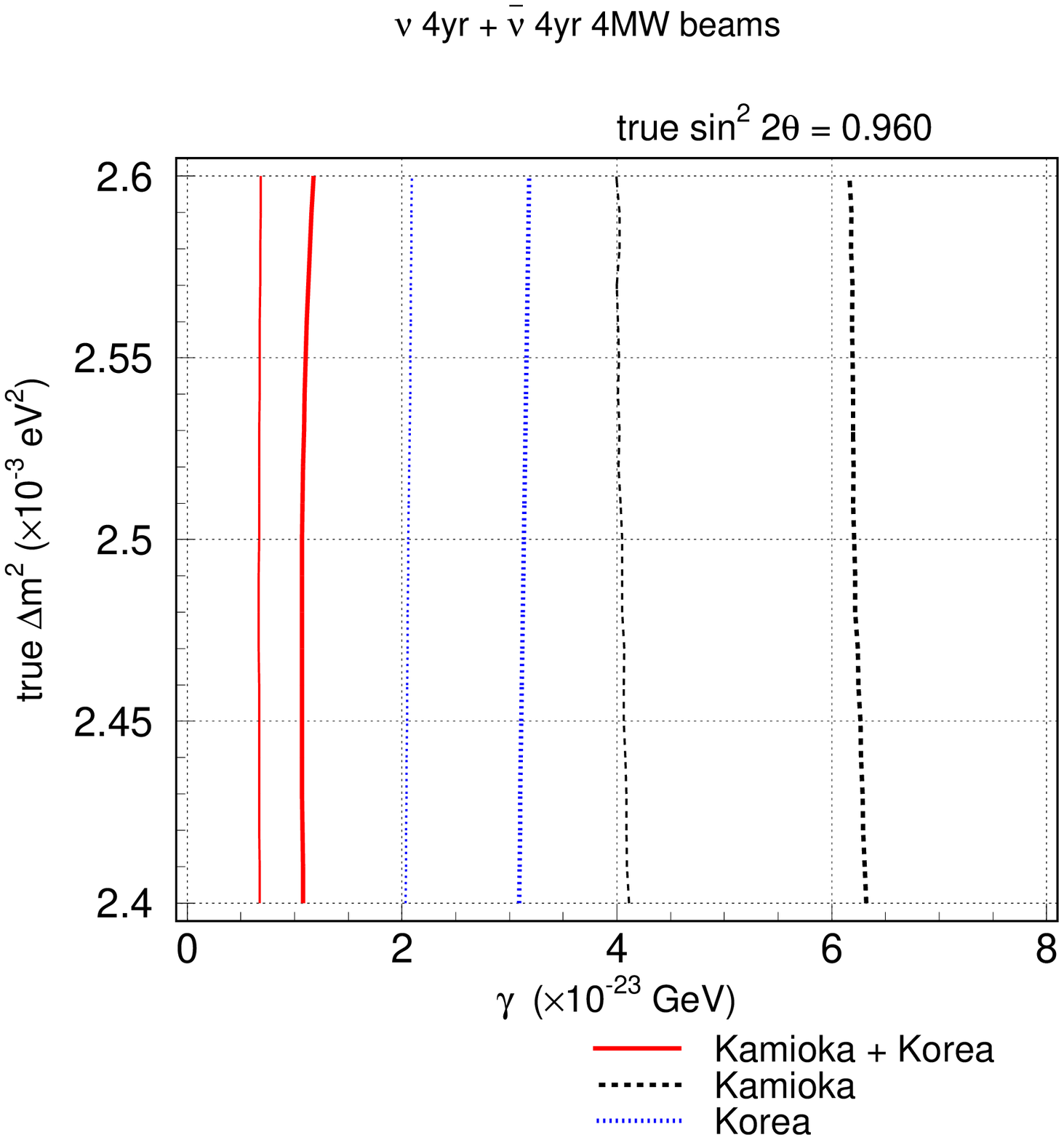} 
\end{center}
\vglue  -0.3cm
\caption{The sensitivity to $\gamma$ as a function of
the true (input) value of $\sin^2 2\theta \equiv \sin^2 2\theta_{23}$ 
(left panel) 
and $\Delta m^2 \equiv \Delta m^2_{32}$ (right panel)
for the case of 1/E dependence of $\gamma(E)$.
The red solid lines are for Kamioka-Korea setting with each 0.27 Mton
detector, while the dashed black (dotted blue) lines are for Kamioka
(Korea) only setting with 0.54 Mton detector.
The thick and the thin lines are for 99\% and 90\%  CL,
respectively.
4 years of neutrino plus 4 years of anti-neutrino running are assumed.
In obtaining the results shown in the left and right panel, 
the input value of $\Delta m_{32}^2$ is taken as 
$+2.5\times 10^{-3}$ eV$^2$ (with positive sign indicating 
the normal mass hierarchy) and that of 
$\sin ^2 2\theta_{23}$ as 0.96, respectively.
}
\label{decoh-gamma-1-over-E}
\end{figure}

First, we examine the case with $\frac{1}{E}$ dependence of $\gamma (E)$.
In Fig.~\ref{sens-examples}, we show the correlations between
$\gamma$ and $\sin^2 2 \theta$ at three experimental setups
for the case where the input values are 
$\gamma=0$, $\Delta m^2=2.5\times 10^{-3}$ eV$^2$ and 
$\sin^2 2 \theta=0.96$.
We immediately find that there are strong correlations between
$\sin^2 2 \theta$ and $\gamma$ for the Kamioka-only and
Korea-only setups. We also
note that the slope of the correlation for the Kamioka-only setup is
different from that for the Korea-only setup. Therefore the Kamioka-Korea
setup can give a stronger bound than each experimental setup.
This advantage can be seen in Fig.~\ref{decoh-gamma-1-over-E},
where we present  the sensitivity regions of $\gamma$ as a function of
$\sin^2 2\theta$ (left panel) and $\Delta m^2$ (right panel).
Note that the sensitivity to $\gamma$ in the Kamioka-Korea
setting is better than the Korea-only and the Kamioka-only settings by
a factor greater than 3 and 6, respectively.

\subsection{Case of energy independent $\gamma (E)$ }

Next, we examine the case of energy independent $\gamma$,
$\gamma (E) = \gamma = constant$.
Repeating the same procedure as before,
we find that the sensitivity to $\gamma$ in the Kamioka-Korea setting
is better than the Korea-only and the Kamioka-only settings by
a factor greater than 3 and 8, respectively, as summarized in
Table \ref{table-gamma}.

\subsection{Case of $E^2$ dependence of $\gamma (E)$ }

Finally, we examine the case with $E^2$ dependence of $\gamma$.
The qualitative features of the sensitivities are similar to
those of the previous two cases. 
The result is that the sensitivity to $\gamma$ in the Kamioka-Korea
setting is better than the Korea-only and the Kamioka-only settings by
a factor greater than 3 and 5, respectively.

\subsection{Comparison between sensitivities of Kamioka-Korea setting
and the existing bound on $\gamma$ }

In Table \ref{table-gamma}, we list, for the purpose of comparison, 
the upper bounds on
$\gamma$ at 90\% CL obtained by analyzing the atmospheric neutrino
data in \cite{QD-bari1} \footnote{We do not quote the bounds
on $\gamma$ obtained from solar and KamLAND neutrinos, since they are
derived from $\nu_e \rightarrow \nu_e$ \cite{fogli07}. In this case,
the neutrino energy is quite low, so that the constraint for $n=-1$
becomes quite strong.}.
We summarize in the table the bounds on $\gamma$ at 90\%  CL
achievable by the Kamioka-only,  the Korea-only and
the Kamioka-Korea settings.  
We use the same ansatz as in \cite{QD-bari1} for parameterizing the
energy dependence of $\gamma (E)$, $n=0, -1$ and 2.

In the case of $\frac{1}{E}$ dependence of $\gamma (E)$, all three
settings can improve the current bound almost by two orders of magnitude.
Note that the best case (Kamioka-Korea) is a factor of 6 better than
the Kamioka-only case.
This case demonstrates clearly that the two-detector setup
is more powerful than the Kamioka-only setup.
We notice that in the case of energy independence of $\gamma (E)$,
Kamioka-Korea two detector setting can improve the current bound by
a factor of $\sim 3$.

In the case of $E^2$ dependence of $\gamma (E)$ the situation
is completely reversed;
The bound imposed by the atmospheric neutrino data surpasses those
of our three settings by almost $\sim$4 orders of magnitude.
Because the spectrum of atmospheric neutrinos spans a wide range
of energy which extends to 100-1000 GeV, it gives much tighter constraints
on the decoherence parameter for quadratic energy dependence of $\gamma (E)$.
In a sense, the current Super-Kamiokande experiment is already a powerful
neutrino spectroscope with a very wide energy range,
and could be sensitive to nonstandard neutrino physics that may affect
higher energy neutrinos such as QD with $\gamma (E) \sim E^2$ or Lorentz
symmetry violation (see Sec.~V for more details).

\begin{table}
\begin{center}
\begin{tabular}{|c|c|c|c|c|}
\hline
Ansats for $\gamma (E)$ & Curent bound (GeV) &
Kamioka-only (GeV)  & Korea-only (GeV) & Kamioka-Korea (GeV)
\\
\hline
$\gamma (E) = \gamma $ (const.) & $< 3.5 \times 10^{-23}$  &
$< 8.7 \times 10^{-23}$   & $< 3.2 \times 10^{-23}$   &
$< 1.1 \times 10^{-23}$
\\
$\gamma (E) = \gamma / E({\rm GeV }) $ & $< 2.0 \times 10^{-21}$  &
$< 4.0 \times 10^{-23} $ & $< 2.0 \times 10^{-23}$  & $ < 0.7 \times 10^{-23}$
\\
$\gamma (E) = \gamma  ( E({\rm GeV}) )^2 $  & $< 0.9 \times 10^{-27}$  &
$< 9.2 \times 10^{-23}$ & $< 6.0 \times 10^{-23}$ & $< 1.7 \times 10^{-23}$
\\
\hline
\end{tabular}
\caption{\label{table-gamma}
Presented are the upper bounds on decoherence parameters
$\gamma$ defined in (\ref{E-dep}) for three possible values of $n$.
The current bounds are based on \cite{QD-bari1} and are at
90\% CL.
The sensitivities obtained by this study are also at 90\% CL
and correspond to the true values of the parameters
$\Delta m^2=2.5 \times 10^{-3} \text{eV}^2$ and
$\sin^2 2\theta_{23} = 0.96$.
}
\end{center}
\end{table}

\section{Violation of Lorentz Invariance}
\label{lorentz}

In the presence of Lorentz symmetry violation by a tiny amount,
neutrinos can have both velocity mixings and the mass mixings, both
are CPT conserving \cite{Coleman1}.
Also there could be CPT-violating interactions in general
\cite{Coleman1,Coleman2,Kostelecky}.
Then, the energy of neutrinos with definite momentum in
ultra-relativistic regime can be written as
\begin{eqnarray}
\frac{m m^{\dagger}} {2p}  = c p + {m^2 \over 2 p} + b,
\label{rel-energy}
\end{eqnarray}
where $m^2$, $c$, and $b$ are $3 \times 3$ hermitian matrices, and the
three terms represent, in order, the effects of velocity mixing,
mass mixing, and CPT violation \cite{Coleman2}.
The energies of neutrinos are eigenvalues of (\ref{rel-energy}),
and the eigenvectors give the ``mass eigenstates''.
Notice that while $c$ is dimensionless quantity, $b$ has
dimension of energy.
For brevity, we will use GeV unit for $b$.

Within the framework just defined above it was shown by
Coleman and Glashow \cite{Coleman2} that
the $\nu_\mu$ disappearance probability in vacuum can be written as
\begin{eqnarray}
P ( \nu_\mu \rightarrow \nu_\mu ) = 1 - \sin^2 2 \Theta~
\sin^2 \left( \Delta L/4 \right)
\label{PLV}
\end{eqnarray}
where the ``mixing anlge'' $\Theta$ and the phase factor $\Delta$ depend
upon seven parameters apart from energy $E$:
\begin{eqnarray}
\Delta \sin 2 \Theta & = & \Delta m^2 \sin 2 \theta_m/E +
2 \delta b e^{i \eta} \sin 2 \theta_b +
2 \delta c e^{i \eta^{'}} E \sin 2 \theta_c, \nonumber
\\
\Delta \cos 2 \Theta & = & \Delta m^2 \cos 2 \theta_m/E +
2 \delta b \cos 2 \theta_b +
2 \delta c  E \cos 2 \theta_c. 
\label{paramLV}
\end{eqnarray}
As easily guessed what is relevant in neutrino oscillation is
the difference in mass squared, and $b$ and $c$
between two mass eigenstates,
$\delta b \equiv b_2 - b_1$ and $\delta c \equiv c_2 - c_1$,
where $c_{i=1,2}$ and $b_{i=1,2}$ are the eigenvalues of the matrix
$c$ and $b$.
The angles $\theta_m$, $\theta_b$ and $\theta_c$ appear in the unitary
matrices which diagonalize the matrices $m^2$, $b$, and $c$, respectively.
There are also two phases $\eta$ and $\eta^{'}$ that cannot be rotated
away by field redefinition.
We work in the convention in which $\cos 2 \theta_m$ and $\cos 2 \theta_b$
are positive, and $\Delta m^2 \equiv m_2^2 - m_1^2$,
$\delta b \equiv b_2 - b_1$ and $\delta c \equiv c_2 - c_1$ can have either
signs.

The survival probability for the anti-neutrino is obtained
by the following substitution:
\begin{eqnarray}
\delta c \rightarrow \delta c , ~~~ \delta b \rightarrow - \delta b
\end{eqnarray}
The difference in the sign changes signify the CPT conserving vs. CPT
violating nature of $c$ and $b$ terms.

The two-flavor oscillation given by (\ref{PLV}) and (\ref{paramLV})
is a too complicated system for full analysis.
Therefore, we make some simplifications in our analysis.
We restrict ourselves into the case
$\theta_m = \theta_b = \theta_c \equiv \theta$ and $\eta = \eta^{'} =0$,
for which one recovers the case treated in \cite{Foot:1998vr}:
\begin{eqnarray}
P ( \nu_\mu \rightarrow \nu_\mu ) = 1 - \sin^2 2 \theta~
\sin^2 \left[ L \left( {\Delta m^2 \over 4 E} + {\delta b \over 2}
+ {\delta c E \over 2} \right) \right],
\label{FLY}
\end{eqnarray}
which still depends on 4 parameters, $\theta$, $\Delta m^2$, $\delta b$
and $\delta c$. One has a similar expression for $\overline{\nu}_\mu$
with $\delta b \rightarrow - \delta b$.
As pointed out in \cite{glashow}, the analysis for violation of
Lorentz invariance with $\delta c$ term is equivalent to testing
the equivalence principle \cite{equivalence}.
The oscillation probability in (\ref{FLY}) looks like the one for
conventional neutrino oscillations due to
$\Delta m^2$, with small corrections due to the Lorentz symmetry violating
$\delta b$ and $\delta c$ terms. In this sense, it may be the most
interesting case to examine as a typical example with the Lorentz symmetry
violation. Note that the sign of $\delta b$ and $\delta c$ can have
different effects on the survival probabilities, so that the bounds on
$\delta b$ and $\delta c$ could depend on their signs, although we will
find that the difference is rather small.

For ease of analysis and simplicity of presentation, we further restrict
our analysis to the case of either $\delta b=0$ and $\delta c \neq 0$,
or $\delta b \neq 0$ and $\delta c=0$.
Notice that the former  is CPT conserving
while the latter is CPT violating.

\subsection{Case with $\delta b=0$ and $\delta c \neq 0$ (CPT conserving)}
\label{CPTC-LV}

\begin{figure}[bhtp]
\begin{center}
\hglue -0.5cm
\includegraphics[width=0.50\textwidth]{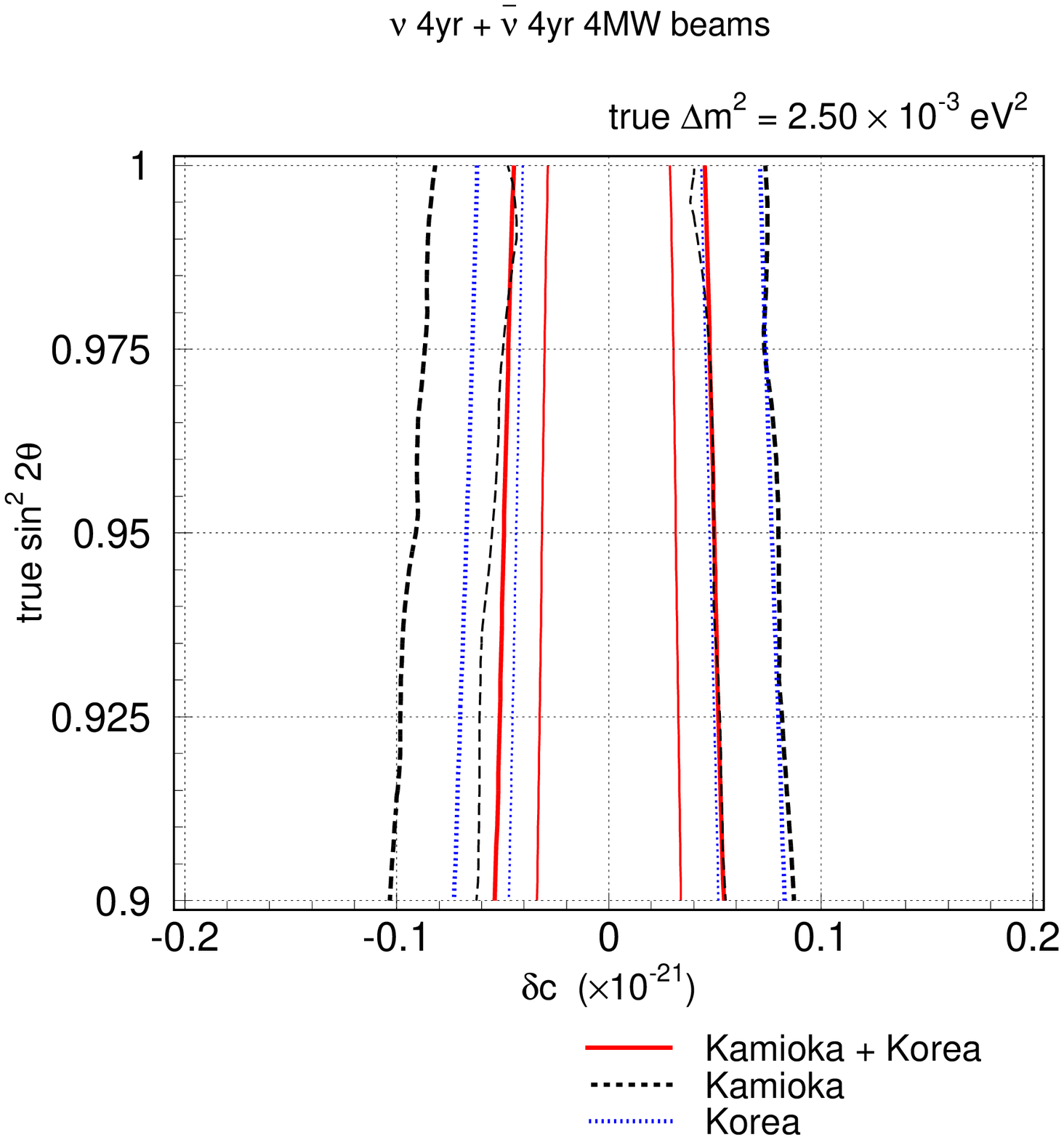}
\hglue  0.3cm
\includegraphics[width=0.50\textwidth]{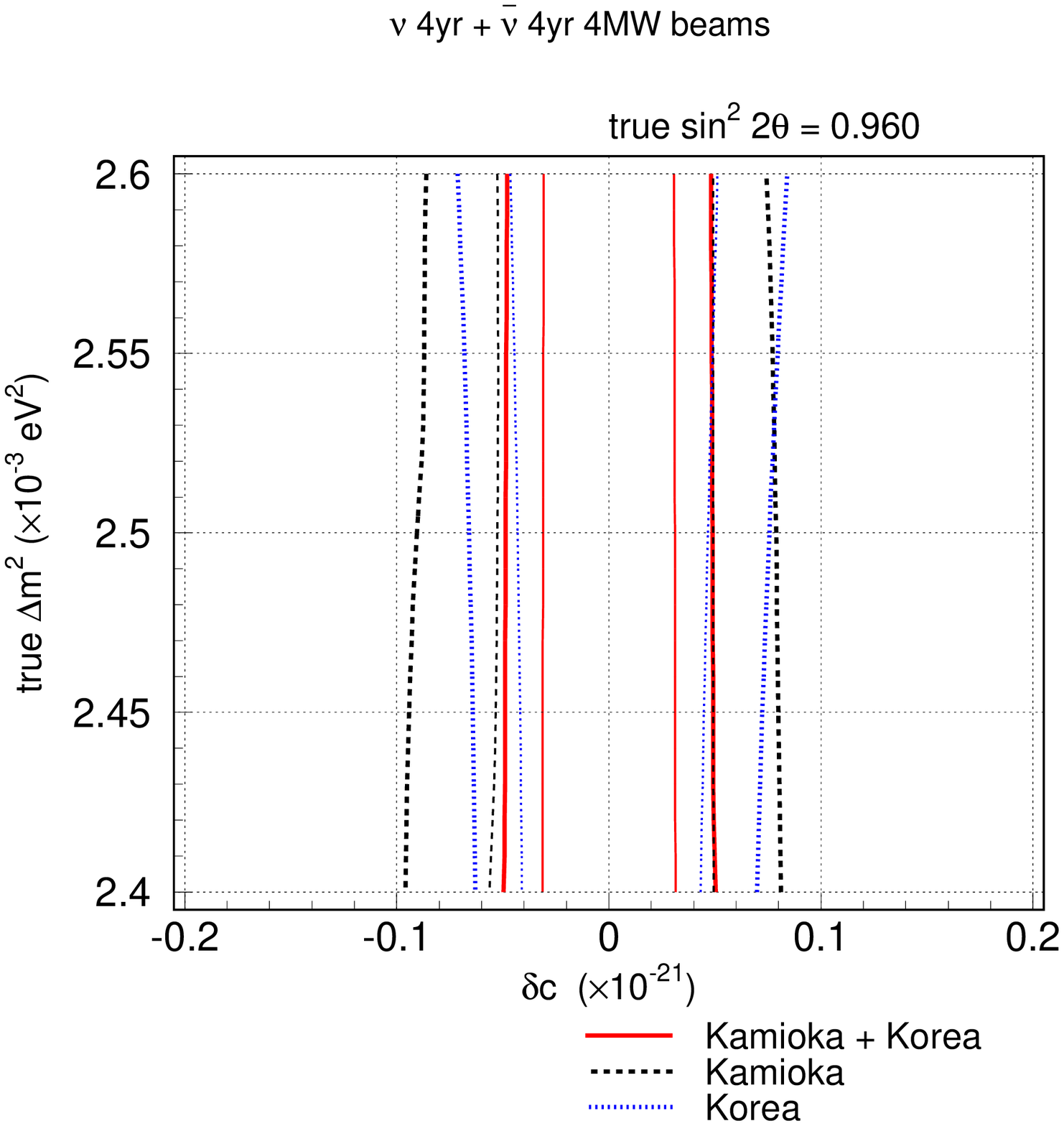}
\end{center}
\vglue  -0.3cm
\caption{The sensitivities to $\delta c$ as a function of
the true (input) value of $\sin^2 2\theta \equiv \sin^2 2\theta_{23}$ 
(left panel) and $\Delta m^2 \equiv \Delta m^2_{32}$ (right panel).
The red solid lines are for Kamioka-Korea setting with each 0.27 Mton
detector,
while the dashed black (dotted blue) lines are for Kamioka (Korea)
only setting with 0.54 Mton detector.
The thick and the thin lines are for 99\% and 90\% CL,
respectively.
4 years of neutrino plus 4 years of anti-neutrino running are assumed.
The other input values of the parameters are identical to those
in Fig.~\ref{decoh-gamma-1-over-E}. }
\label{lorentz-v-b0}
\end{figure}

We first examine violation of Lorentz invariance
in the case of $\delta b=0$ and $\delta c \neq 0$.
In Fig.~\ref{lorentz-v-b0} we present the region of
allowed values of $\delta c$ as a function of
$\sin^2 2\theta$ (left panel) and $\Delta m^2$ (right panel).
Unlike the case of quantum decoherence,
the sensitivities to $\delta c$ achieved by the Kamioka-Korea
setting is slightly better than those of the Korea-only and the
Kamioka-only settings but not so much.
The sensitivity is weakly correlated to $\theta$, and the best sensitivity
is achieved at the maximal $\theta$.
There is almost no correlation to $\Delta m^2$.

\subsection{Case with $\delta c=0$ and $\delta b \neq 0$ (CPT violating)}
\label{CPTV-LV}

\begin{figure}[bhtp]
\begin{center}
\hglue -0.5cm
\includegraphics[width=0.50\textwidth]{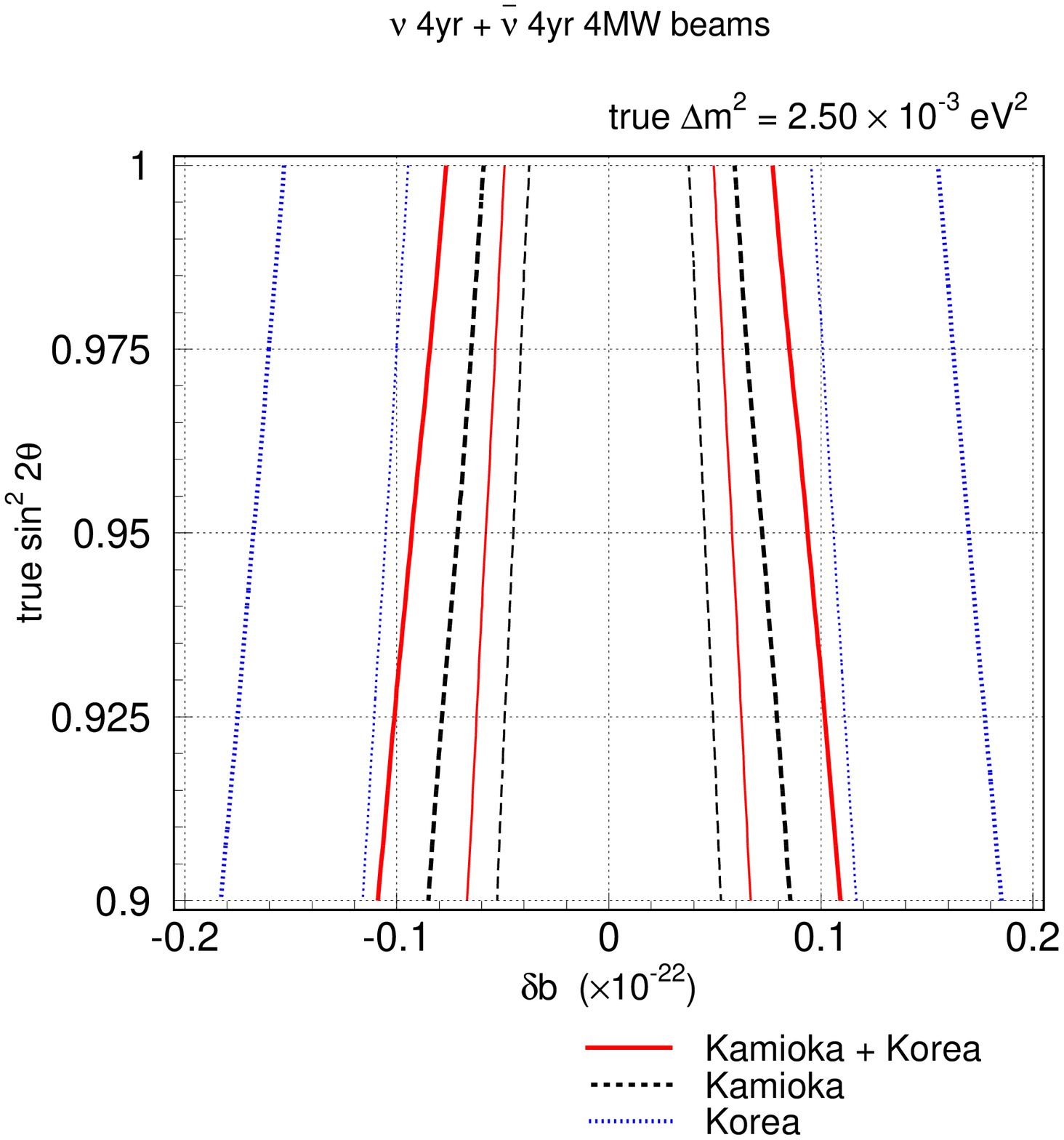}
\hglue  0.3cm
\includegraphics[width=0.50\textwidth]{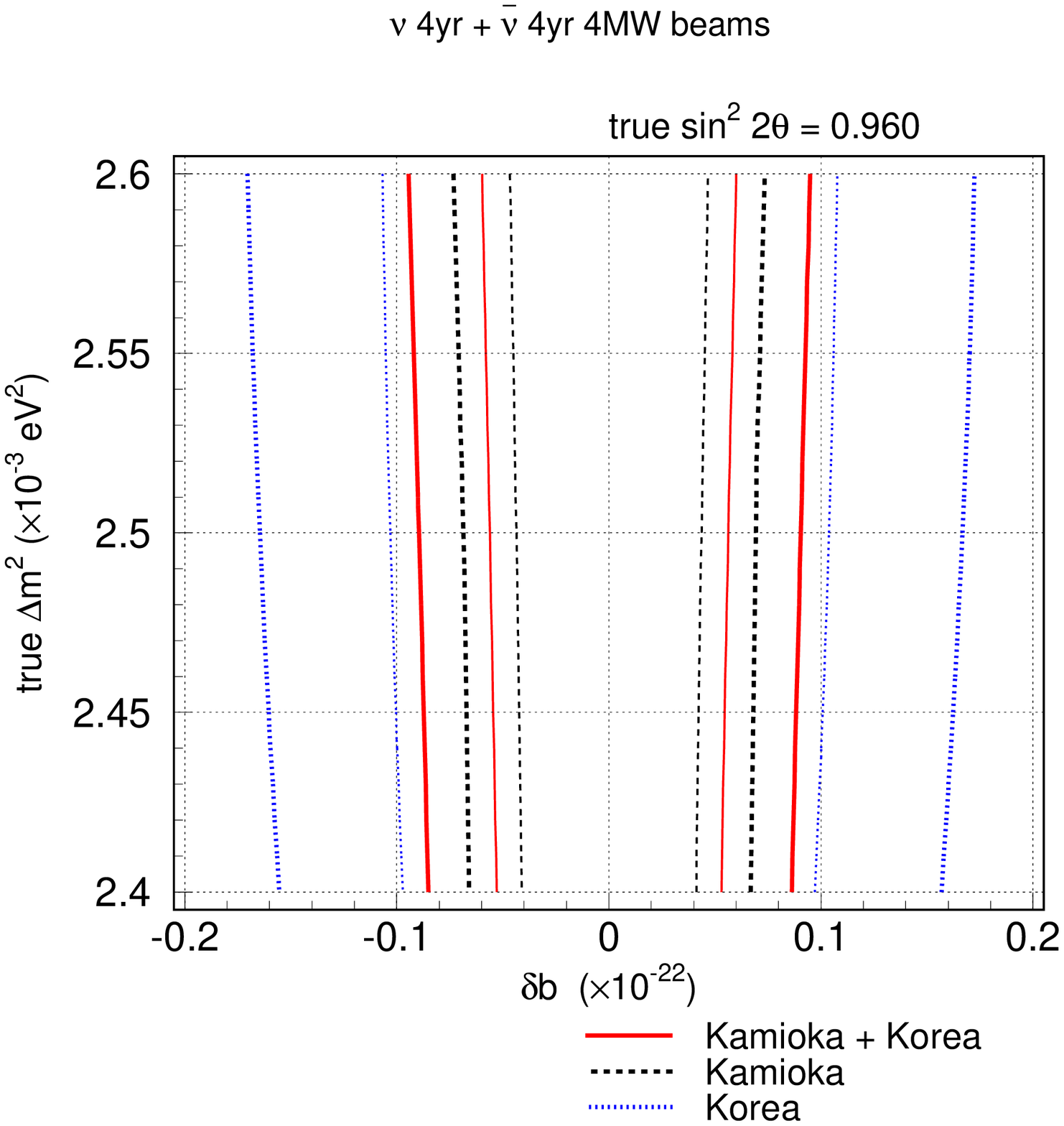}
\end{center}
\vglue  -0.3cm
\caption{The same as in Fig.~\ref{lorentz-v-b0} but for the
case of non-vanishing $\delta b$.
}
\label{lorentz-v-c0}
\end{figure}

The allowed regions with violation of Lorentz invariance
in the case of $\delta c=0$ and $\delta b \neq 0$ presented in
Fig.~\ref{lorentz-v-c0} have several unique features.
First of all, unlike the system with decoherence,
the sensitivity is greatest in the Kamioka-only setting,
though the one by the Kamioka-Korea setting is only slightly less
by about $15-20$\%.
Whereas, the sensitivity by the Korea-only setting is much worse,
more than a factor of 2 compared to the Kamioka-only setting.
The reason for this lies in the $\nu_\mu$ and $\overline{\nu}_\mu$
survival probabilities.
In this scenario, the effect of the non-vanishing $\delta b$
appears as the difference in the oscillation frequency between
neutrinos and anti-neutrinos, if the energy dependence is neglected.
In this case, the measurement at different baseline is not very
important.
Then the Kamioka-only setup turns out to be
slightly better than the Kamioka-Korea setup.
This case is also unique by having the worst sensitivity at the
largest value of $\Delta m^2$ (right panel).
Also, the correlation of sensitivity to $\sin^2 2\theta$ (left panel)
is strongest among the cases examined in this paper,
with maximal sensitivity at maximal $\theta$.

\subsection{Comparison between sensitivities of Kamioka-Korea setting and
the existing  bounds }
\label{SUMMARY-LV}

We summarize the results of the previous subsections in 
Table~\ref{table-liv}, along
with the present bounds on $\delta c$ and $\delta b$, respectively.
We quote the current bounds on $\delta c$'s from 
Ref.s~\cite{Fogli:1999fs,macro_LV}
which was obtained by the atmospheric neutrino data, 
\begin{equation}
|\delta c_{\mu \tau}| \lsim 3\times 10^{-26}. 
\end{equation}
We note that the current bound on $\delta c_{\mu \tau}$ obtained by 
atmospheric neutrino data is quite strong. 
The reason why the atmospheric neutrino data give much stronger 
limit is that the relevant energy is much higher (typically $\sim 100$ GeV) 
than the one we are considering ($\sim$ 1 GeV) and 
the baseline is larger, as large as the Earth diameter.

For the bound on $\delta b$,  Barger et al. \cite{barger} argue that
\begin{equation}
|\delta b_{\mu\tau}| < 3 \times 10^{-20}~{\rm GeV}
\end{equation}
from the analysis of the atmospheric neutrino data.
Let us compare the sensitivity on $\delta b$ within our two detector setup
with the sensitivity at a neutrino factory.  Barger et al. \cite{barger}
considered a neutrino factory with $10^{19}$ stored muons
with 20 GeV energy, and 10 kton detector, and concluded that it can probe
$|\delta b | < 3 \times 10^{-23}$ GeV. The Kamioka-Korea two detector setup
and Kamioka-only setup have five and six times better sensitivities
compared with the neutrino factory with the assumed configuration.
Of course the sensitivity of a neutrino factories could be improved
with a larger number of stored muons and a larger detector.
A more meaningful comparison would be possible, only when one has
configurations for both experiments which are optimized for the purposes
of each experiment. Still we can conclude that the two-detector setup
could be powerful to probe the Lorentz symmetry violation.

\begin{table}
\begin{center}
\begin{tabular}{|c|c|c|c|c|}
\hline
LV parameters & Curent bound &
Kamioka-only  & Korea-only &  Kamioka-Korea
\\
\hline
$| \delta c | $ & $ \lesssim 3\times 10^{-26}$  &
$\lesssim 5 \times 10^{-23}$   & $\lesssim 4 \times 10^{-23}$   &
$\lesssim 3 \times 10^{-23}$
\\
$| \delta b |$ (GeV) & $ < 3.0 \times 10^{-20}$  &
$\lesssim 1 \times 10^{-23} $ & $\lesssim 0.5 \times 10^{-23}$  &
$\lesssim 0.6 \times 10^{-23}$
\\
\hline
\end{tabular}
\caption{\label{table-liv}
Presented are the upper bounds on the velocity mixing parameter
$\delta c$ and the CPT violating parameter $\delta b$ (in GeV)
for the case where $\theta_m = \theta_b = \theta_c \equiv \theta$ and 
$\eta = \eta^{'} =0$. 
The current bounds are based on \cite{Fogli:1999fs,barger} and are at
90\% CL.
The sensitivities obtained in this study are also at 90\% CL
and correspond to the true values of the parameters
$\Delta m^2=2.5 \times 10^{-3} \text{eV}^2$ and
$\sin^2 2\theta_{23} = 0.96$.
}
\end{center}
\end{table}


\section{Nonstandard neutrino interactions with matter}
\label{NSI}

It was suggested that neutrinos might have nonstandard
neutral current interactions with matter 
\cite{wolfenstein,grossmann,berezhiani,NSI},
$\nu_\alpha + f \rightarrow \nu_\beta + f$
($\alpha, \beta = e, \mu , \tau$), with $f$ being the up quarks,
the down quarks and electrons.
This effect may be described by a low energy effective
Hamiltonian for new nonstandard interactions (NSI) of neutrinos:
\begin{eqnarray}
H_{\rm NSI} = 2 \sqrt{2} G_F \left( \bar{\nu}_\alpha \gamma_\rho \nu_\beta
\right)~\left( \varepsilon_{\alpha\beta}^{f f^{'} L} \bar{f}_L \gamma^\rho f_L^{'}
+ \varepsilon_{\alpha \beta}^{f f^{'} R} \bar{f}_R \gamma^\rho f^{'}_R \right)
+ H.c.
\end{eqnarray}
where $\varepsilon_{\alpha\beta}^f \equiv \varepsilon_{\alpha\beta}^{f L} +
\varepsilon_{\alpha\beta}^{f R}$ and
$\varepsilon_{\alpha\beta}^{f P} \equiv \varepsilon_{\alpha\beta}^{f f P}$.
It is known that the presence of such NSI can 
affect production and/or detection processes of neutrinos 
as well as propagation of neutrinos in matter. 
In this work, for simplicity, we consider the impact of NSI 
only for propagation.
By using $\varepsilon_{\alpha \beta}$ defined as
$\varepsilon_{\alpha\beta} \equiv \sum_{f=u,d,e} \varepsilon^{f}_{\alpha\beta}
n_f / n_e $,
the effects of NSI may be summarized by a term with dimensionless
parameters $\varepsilon_{\alpha\beta}$ in  the effective Hamiltonian
\begin{eqnarray}
  \label{eq:neweffH}
  H_{\rm eff} = \sqrt{2} G_F N_e \left( \begin{array}{ccc}
      \varepsilon_{ee}     & \varepsilon_{e\mu} & \varepsilon_{e\tau} \\
      \varepsilon_{e \mu}^\ast  & \varepsilon_{\mu\mu}  & \varepsilon_{\mu\tau} \\
      \varepsilon_{e \tau}^\ast 
& \varepsilon_{\mu\tau}^\ast & \varepsilon_{\tau\tau}
               \end{array}
               \right)
\label{NSI-hamiltonian}
\end{eqnarray}
which is to be added to the standard matter term
$\sqrt{2} G_F N_e \text{ diag.}(1, 0, 0) $  \cite{wolfenstein} in the
evolution equation of neutrinos.
Here, $G_F$ is the Fermi constant, $N_e$ denotes the averaged
electron number density along the neutrino trajectory in
the earth.
The existing constraints on $\varepsilon_{\alpha\beta}$ 
are worked out in 
\cite{constraint_nsi,constraint_nsi_lep,constraint_nsi_atm}:
\begin{equation}
\left( \begin{array}{ccc}
-0.9 < \varepsilon_{ee} < 0.75  &
| \varepsilon_{e\mu} |  \lesssim  3.8 \times 10^{-4}  &
| \varepsilon_{e\tau} | \lesssim 0.25
\\
&  -0.05 < \varepsilon_{\mu\mu} < 0.08  &
 | \varepsilon_{\mu\tau} | \lesssim  0.15
\\
&&  | \varepsilon_{\tau\tau} | \lesssim 0.4
\end{array}
\right)\ .
\end{equation}
Note that the bounds on $\varepsilon_{\mu\tau}$ and 
$\varepsilon_{\tau\tau}$
are coming from atmospheric neutrino data~\cite{constraint_nsi_atm}
and LEP data~\cite{constraint_nsi_lep}, respectively, 
which are relatively weak 
and we wish to investigate how much we can improve
these bounds at the Kamioka-Korea two detector setup.

In this work we truncate the system so that we confine into
the $\mu - \tau$ sector of the neutrino evolution.
Then, the time evolution of the neutrinos in flavor basis can be
written as
\begin{eqnarray}
  \label{eq:evol}
i {d\over dt} \left( \begin{array}{c}
                   \nu_\mu \\ \nu_\tau
                   \end{array}  \right)
 = \left[ U \left( \begin{array}{cc}
                   0  & 0  \\
                   0   & \frac{ \Delta m^2_{32} }{2 E}
                   \end{array} \right)
            U^{\dagger} +  a \left( \begin{array}{cc}
            0 & \varepsilon_{\mu\tau} \\
             \varepsilon_{\mu\tau}^\ast & \varepsilon_{\tau\tau} -
                \varepsilon_{\mu\mu}
                   \end{array}
                   \right) \right] ~
\left( \begin{array}{c}
                    \nu_\mu \\ \nu_\tau
                   \end{array}  \right),
\end{eqnarray}
where $U$ is the flavor mixing matrix and $a \equiv \sqrt{2} G_F N_e$.
In the 2-2 element of the NSI term in the Hamiltonian is of the form
$\varepsilon_{\tau \tau} - \varepsilon_{\mu\mu}$ because the oscillation
probability depend upon $\varepsilon$'s only through this combination.
The evolution equation for the anti-neutrinos are given by changing the signs
of $a$ and replacing $U$ by $U^*$.

In fact, one can show that the truncation to the $2 \times 2$
sub system is a good approximation.
In the full three flavor framework the $\nu_{\mu}$ disappearance
oscillation probability can be computed to leading order of NSI
as \cite{KLOS}
\begin{eqnarray}
 P( \nu_{\mu} \rightarrow \nu_{\mu} ) &=&
    1 - \sin^2 2 \theta_{23}  \sin^{2} \Delta_{32}
    \nonumber\\
   &-&  |\varepsilon_{\mu\tau}| \cos\phi_{\mu\tau}  \sin 2 \theta_{23}
    \left( a L \right)
    \left[
        \sin^2 2 \theta_{23}  \sin 2 \Delta_{32}
       +  \cos^2 2 \theta_{23}
         \frac{2 }{ \Delta_{32} } \sin^{2} \Delta_{32}
     \right]
     \nonumber\\
   & -& \frac{1}{2}\left( \varepsilon_{\tau\tau} - \varepsilon_{\mu\mu} \right)
     \sin^2 2 \theta_{23}  \cos 2 \theta_{23}
    \left( a L \right)
    \left[
        \sin 2 \Delta_{32}
       - \frac{ 2 }{ \Delta_{32} } \sin^{2} \Delta_{32}
     \right]
     \nonumber\\
      &+& \mathcal{O}\Big( \frac{\Delta m^2_{21}}{\Delta m^2_{31}} \Big)
      + \mathcal{O} ( s_{13} )
      + \mathcal{O} ( \varepsilon^2 ),
  \label{eq:Pmumu}
\end{eqnarray}
where
$\Delta_{32} \equiv \frac{ \Delta m^2_{32} L}{4E}$ and
$\phi_{\mu\tau} $ is the phase of $\varepsilon_{\mu\tau}$.
The result in  (\ref{eq:Pmumu}) indicates that the truncation is legitimate
if $\varepsilon$'s are sufficiently small.
Also note that the dependence on $\varepsilon_{\tau\tau} -
                                  \varepsilon_{\mu\mu}$
goes away for $\sin^2 \theta = 0.5$, so that the muon disappearance becomes
insensitive to this combination of $\varepsilon$'s. We will find that
the sensitivity on $\varepsilon_{\tau\tau} - \varepsilon_{\mu\mu}$ strongly
depends on $\sin^2 \theta$ ($\theta$ being maximal or not) for this reason.
In the following we set $\epsilon_{\mu\mu} = 0$ so that the sensitivity  
contours presented for $\epsilon_{\tau\tau}$ actually means those for   
$\epsilon_{\tau\tau} - \epsilon_{\mu\mu}$.
Moreover, for simplicity, we assume that $\epsilon_{\mu\tau}$ is a real
by ignoring its phase. 


\begin{figure}[bhtp]
\begin{center}
\includegraphics[height=18cm]{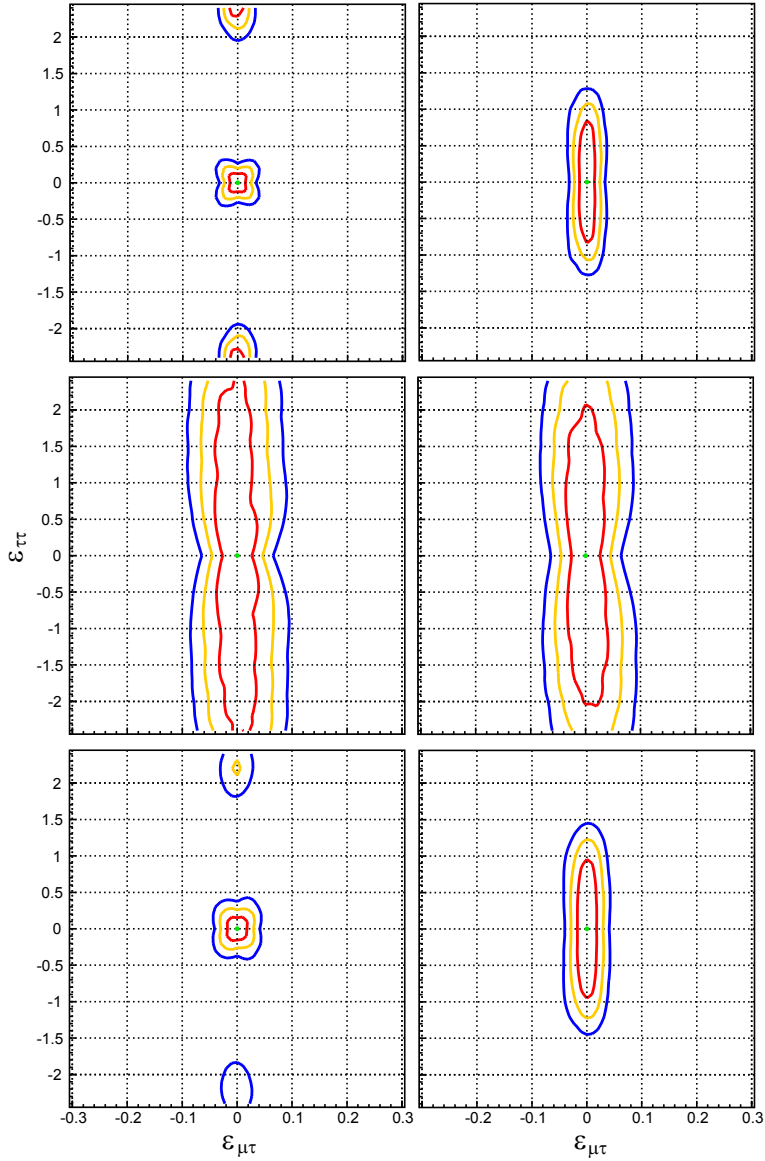}
\end{center}
\vglue -2cm
\caption{
The allowed regions in
$\varepsilon_{\mu\tau} - \varepsilon_{\tau\tau}$ space for 4 years neutrino
and 4 years anti-neutrino running.
The upper, the middle, and the bottom three panels are for
the Kamioka-only setting, the Korea-only setting, and the
Kamioka-Korea setting,  respectively.
The left and the right panels are for cases with
$\sin^2 \theta \equiv \sin^2 \theta_{23} = 0.45$ and 0.5, respectively.
The red, the yellow, and the blue lines indicate the allowed regions
at 1$\sigma$, 2$\sigma$, and 3$\sigma$ CL, respectively.
The input value of $\Delta m^2_{32}$ is taken as $2.5\times 10^{-3}$ eV$^2$. 
}
\label{sens-kam_kor}
\end{figure}

In Fig.~\ref{sens-kam_kor}, presented are the allowed regions in
$\varepsilon_{\mu\tau} - \varepsilon_{\tau \tau}$
space for 4 years neutrino and 4 years anti-neutrino running of the
Kamioka-only  (upper panels),  the Korea-only (middle panels),
and the Kamioka-Korea (bottom panels) settings.
The input values $\varepsilon_{\mu\tau}$ and $\varepsilon_{\tau\tau}$
are taken to be vanishing.

As in the CPT-Lorentz violating case studied in Sec.~\ref{CPTV-LV}
and unlike the system with decoherence, the Korea-only setting gives
much worse sensitivity compared to the other two settings.
Again the Kamioka-only setting has a slightly better sensitivity than the
Kamioka-Korea setting.
However we notice that the Kamioka-only setting has multiple
 $\varepsilon_{\tau\tau}$ solutions for $\sin^2 \theta_{23} = 0.45$.
The fake solutions are nearly eliminated in the Kamioka-Korea setting.

The sensitivities of three experimental setups at 2 $\sigma$ CL 
can be read off from
Fig.~\ref{sens-kam_kor}. The approximate
2 $\sigma$ CL sensitivities of the
Kamioka-Korea setup  for
$\sin^2 \theta = 0.45 ~(\sin^2 \theta = 0.5)$ are:
\begin{equation}
| \epsilon_{\mu\tau} |  \lesssim  0.03 ~(0.03),~~~
| \epsilon_{\tau\tau}|  \lesssim  0.3 ~(1.2).
\end{equation}
Here we neglected a barely allowed region near $|\epsilon_{\tau\tau}| = 2.3$,
which is already excluded by the current data. 
Note that the sensitivity on $\varepsilon_{\tau\tau}$ becomes weak for
maximal mixing ($\sin^2 \theta =0.5$), for the above mentioned reason
(see Eq.~(\ref{eq:Pmumu}) and the subsequent discussions).
The Kamioka-Korea setup can improve the current
bound on $| \epsilon_{\mu\tau} |$ 
by factors of $\sim$ 5 whereas 
the bound on $| \epsilon_{\tau\tau}| $ we obtained is 
comparable to (worse than) the current bound
for $\sin^2 \theta = 0.45 ~(\sin^2 \theta = 0.5)$. 
A similar statement applies to the case for the Kamioka-only setup.

There are a large number of references which studied the effects of
NSI and the sensitivity reach to NSI by the ongoing and the various
future projects.
We quote here only the most recent ones which focused on
sensitivities by superbeam and reactor experiments~\cite{KLOS}
and neutrino factory~\cite{CMNUZ}.
The earlier references can be traced back through the bibliography of
these papers.

By combining future superbeam experiment, T2K~\cite{T2K} and 
reactor one, Double-Chooz \cite{reactor13}, the authors of \cite{KLOS}
obtained the sensitivity of $|\epsilon_{\mu \tau}|$ to be $\sim$ 0.25
when it is assumed to be real (no CP phase) while essentially 
no sensitivity to $\epsilon_{\tau \tau}$ is expected. 
The same authors also consider the case of NO$\nu$A experiment~\cite{NOVA} 
combined with some future upgraded reactor experiment with larger detector 
as considered, e.g., in \cite{DC200,angra} and obtained 
$\epsilon_{\mu \tau}$ sensitivity of about 0.05 which is 
comparable to what we obtained. 

While essentially no sensitivity of $\epsilon_{\tau \tau}$ is expected
by superbeam, future neutrino factory with the so called golden channel 
$\nu_e \to \nu_\mu$ and $\bar{\nu}_e \to \bar{\nu}_\mu$, 
could reach the sensitivity to $\epsilon_{\tau \tau}$ at the level of 
$\sim$ 0.1-0.2~\cite{CMNUZ}. 
Despite that the sensitivity to $\epsilon_{\mu \tau}$ by neutrino factory 
was not derived in ~\cite{CMNUZ}, from Fig. 1 of this reference, 
one can naively expect that the sensitivity to $\epsilon_{\mu \tau}$
is similar to that of $\epsilon_{ee}$ which is  $\sim$ 0.1 or so.
We conclude that the sensitivity we obtained for 
$\epsilon_{\mu \tau}$ is not bad. 

\section{Concluding remarks}
\label{conclusion}

The Kamioka-Korea two detector system was shown to be a powerful
setup for lifting neutrino parameter degeneracies and probing
CP violation in neutrino oscillation.
In this paper, we study sensitivities of this setup
to nonstandard neutrino physics such as quantum decoherence,
tiny violation of Lorentz symmetry, and nonstandard interactions of
neutrinos with matter. In most cases, two detector setup is more sensitive
than a single detector at Kamioka or Korea, except for the Lorentz violation
with $\delta b \neq 0$, and the nonstandard neutrino interactions with
matter.
The sensitivities of three experimental setups at 90\% CL are
summarized in Table \ref{table-gamma} and
Table \ref{table-liv} for quantum decoherence and Lorentz
symmetry violation with/without CPT symmetry, respectively.
We can say that future long baseline experiments with two detector  
setup can
improve the sensitivities on nonstandard neutrino physics in many cases.
We believe that it is a useful addition to the physics capabilities  
of the Kamioka-Korea two-detector setting that are already demonstrated, 
namely, resolution  
of the mass hierarchy and resolving CP and the octant degeneracies.

\begin{acknowledgments}
  One of the authors (P.K.) is grateful to ICRR where a part of
  this research has been performed.
  Two of us (H.M. and H.N.) thank Theoretical Physics Department of
  Fermi National Accelerator Laboratory for hospitalities extended to
  them in the summer of 2007.
  This work was supported in part by KAKENHI,
  the Grant-in-Aid for Scientific Research,
  No. 19340062, Japan Society for the Promotion of Science,
  by Funda\c{c}\~ao de Amparo \`a Pesquisa do Estado de Rio de
  Janeiro (FAPERJ) and Conselho Nacional de Ci\^encia e Tecnologia
  (CNPq), and
  by KOSEF through CHEP at Kyungpook National University.
\end{acknowledgments}

\end{document}